# Epitaxial stabilization of metastable 3C BaRuO$_3$ thin film with ferromagnetic non-Fermi liquid phase

Sang A Lee[1,2], Jong Mok Ok[3], Jegon Lee[1], Jae-Yeol Hwang[2], Sangmon Yoon[3], Se-Jeong Park[4], Sehwan Song[5], Jong-Seong Bae[6], Sungkyun Park[5], Ho Nyung Lee[3], and Woo Seok Choi[1*]

[1]Department of Physics, Sungkyunkwan University, Suwon 16419, Korea

[2]Department of Physics, Pukyong National University, Busan 48513, Korea

[3]Materials Science and Technology Division, Oak Ridge National Laboratory, Oak Ridge, TN 37831, U.S.A.

[4]Application Group, Korea I.T.S. Co. Ltd., Seoul 06373, Korea

[5]Department of Physics, Pusan National University, Busan 46241, Korea

[6]Busan Center, Korea Basic Science Institute, Busan 46742, Korea

* e-mail: choiws@skku.edu

# Abstract

Thin films of perovskite Ruthenates of the general formula $A$RuO$_3$ ($A$ = Ca and Sr) are versatile electrical conductors for viable oxide electronics. They are also scientifically intriguing, as they exhibit non-trivial electromagnetic ground states depending on the $A$-site element. Among them, realization of the cubic perovskite (3C) BaRuO$_3$ thin film has been a challenge so far, because the 3C phase is metastable with the largest formation energy among the various polymorph phases of BaRuO$_3$. In this study, we successfully prepared 3C BaRuO$_3$ thin films employing epitaxial stabilization. The 3C BaRuO$_3$ thin films show itinerant ferromagnetism with a transition temperature of ~48 K and a non-Fermi liquid phase. The epitaxial stabilization of the 3C BaRuO$_3$ further enabled us to make a standard comparison of perovskite Ruthenates thin films, thereby establishing the importance of the Ru-O orbital hybridization in understanding the itinerant magnetic system.

# Introduction

The epitaxial stabilization of transition metal oxides grants us access to metastable structural phases with fascinating emergent properties. Thermodynamically, epitaxial stabilization is determined by the competition between the surface diffusion and nucleation energy during thin film growth.[1] Therefore, by modifying the surface of a crystalline structure,[2] it is possible to stabilize different structural phases of polymorphs with small differences in their formation energy. As a prominent example, multiferroic $R$MnO$_3$ ($R$ = rare earth elements) thin films were epitaxially stabilized in both the stable orthorhombic and metastable hexagonal phases by utilizing the surface crystalline symmetries of the substrates.[3, 4] In addition, owing to the small lattice mismatch, a TiO$_2$ thin film on a SrTiO$_3$ (STO) (001) substrate was stabilized in the anatase phase, which is not thermodynamically stable in bulk.[5] This approach has been widely employed to grow various oxide thin films, including SrMnO$_3$, BaMnO$_3$, and CaCuO$_2$.[6-8]

Interestingly, there are also cases in which the epitaxial stabilization of metastable structures in thin-film form is possible even when a fully strained state is not achieved, including domain matching epitaxy in oxides such as B-phase VO$_2$ thin film on STO substrate.[9-11] Even without the coherent chemical bonding at the interface owing to a large lattice mismatch, a relaxed thin film would still be able to maintain the same structural surface symmetry of the substrate. Several examples, including TiO$_2$/YSZ, LiNbO$_3$/STO, Sr$_2$RuO$_4$/LaAlO$_3$ (LAO) (thin film/substrate) systems have been reported.[5, 12, 13]

Within this context, BaRuO$_3$ (BRO) is an excellent model system for the verification of symmetry stabilization because it possesses various polymorph phases with small differences in their formation energy. The most stable BRO phase is a nine-layered rhombohedral (9R, space group: $R$-$3m$) structure with three face-sharing RuO$_6$ octahedra connected by corner-sharing along the out-of-plane direction.[14] When bulk material is synthesized at pressure greater than 3 GPa, the structure of BRO becomes four-layered hexagonal (4H), six-layered hexagonal (6H), or cubic perovskite (3C).[15, 16] The different crystal structures of BRO induce substantially distinct electronic and magnetic ground states. BRO with 9R, 4H, or 6H phase is known to be a paramagnetic metal with short-range antiferromagnetic fluctuations,[17] whereas 3C BRO, which consists solely of three-dimensional corner-shared $RuO_6$ octahedra, is known to be a ferromagnetic metal.[16] So far, most studies on the BRO polymorphs focused on bulk samples. The fabrication of epitaxially stabilized 3C BRO thin film has been known to be challenging because of the natural formation of polycrystalline and/or mixed polymorph phases. A systematic study of 3C BRO using epitaxial stabilization would facilitate our understanding and ensure the controllability of the intriguing polymorph system. Moreover, the realization of 3C BRO in thin film form would enable a systematic investigation of the $4d$ perovskite Ruthenate thin films ($ARuO_3$, $A$ = Ca, Sr, and Ba) with non-trivial electromagnetic ground states that depend on the $A$-site ion.

In this study, we successfully realized epitaxially-stabilized ferromagnetic metallic 3C BRO thin films using pulsed laser deposition (PLD). We present the structural, optical, electric, and magnetic properties of 3C BRO to complement existing knowledge of the perovskite Ruthenate family comprising of $CaRuO_3$ (CRO), $SrRuO_3$ (SRO), and BRO. Not only did the 3C BRO thin film exhibited a clear ferromagnetic hysteresis with a transition temperature of ~48 K, but its behavior resembled that of a non-Fermi liquid (NFL). A systematic comparison with the other perovskite Ruthenates led to a correlation between the $p$–$d$ hybridization strength and the ferromagnetic ordering of the spins.

## Results and Discussions

To verify the idea of symmetry stabilization in cubic perovskites, we deposited BRO thin films on an STO (001) substrates with the square surface symmetry as an extension of "cube-on-cube" epitaxy (**Figure 1**). We controlled the growth parameters of PLD to optimize the stabilization, as shown in Figure 1(a). Notably, by fine-tuning the oxygen partial pressure ($P(O_2)$) and temperature ($T_g$) during film growth, we obtained BRO thin films with various polymorph crystalline phases. Particularly, the $P(O_2)$ largely modified the plume dynamics during growth, resulting in dramatic changes in the resultant thin films.[18, 19] The growth parameter mapping of the polymorph BRO thin films on STO substrates (Figure 1(a)) indicates that at high $P(O_2)$ (> 10 mTorr), the thin films have multiple phases that are not affected by the surface symmetry of the substrate. Depending on $T_g$, we not only succeeded in obtaining the 9R phase, which is the most stable phase in bulk, but also the 6H and 4H phases. On the contrary, the 3C BRO thin films were preferentially stabilized when grown at low $P(O_2)$ (≤ 10 mTorr). The 3C BRO films were stable across a wide range of $T_g$, namely, 650°C ≤ $T_g$ < 750°C. The region marked in yellow on the growth parameter map in Figure 1(a) represents the growth window of 3C BRO. Figure 1(b) shows the X-ray diffraction (XRD) $\theta$–$2\theta$ scans for BRO thin films grown at $P(O_2)$ = 300 and 1 mTorr (indicated by two dotted circles in Figure 1(a)). Detailed peak indexing was possible for the BRO thin films. The thin film grown at 300 mTorr showed multiple polymorph phases including 4H and 9R, whereas that grown at 1 mTorr showed the symmetry-stabilized phase-pure 3C phase. Figure S1 summarizes the $P(O_2)$ dependence of the BRO thin films grown at 750 °C in greater detail, demonstrating the effectiveness of the epitaxial stabilization and the high quality of our thin films.

To further clarify the epitaxial stabilization of 3C BRO thin films on STO substrates, we performed XRD $\varphi$-scan and reciprocal space mapping (RSM), as shown in Figure 1(c) and 1(d), respectively, for the 3C BRO thin film grown at $P(O_2)$ = 1 mTorr. The $\varphi$-scan of 3C BRO (200) and STO (200) in-plane reflections revealed identical four-fold symmetry peaks indicating extended "cube-on-cube" epitaxy of the perovskite BRO (001) thin film on the STO (001) substrate. Notably, all the other polymorphs of BRO would have resulted in three-fold or six-fold symmetries.[20] The RSM around the (103) Bragg reflection of the STO substrate indicates that the 3C BRO thin film is not fully strained. The in-plane $\theta$–$2\theta$ scan around the (200) Bragg reflection (Figure S2(a)) and the wide-range RSM around the (204) Bragg reflection of the STO substrate further support the formation of single phase 3C BRO with strain relaxation.[21-23] The experimental lattice parameters of the 3C BRO thin film obtained from our XRD analyses were 4.16 and 3.99 Å, along the out-of-plane and in-plane directions, respectively (tetragonality $c/a$ = 1.03). Because of the relaxation of the epitaxial strain, the out-of-plane lattice parameter is larger than the in-plane one and this stabilizes the tetragonal symmetry of the thin film. This indicates that the epitaxial strain, which is necessary for the achievement of epitaxial stabilization, partially remains despite of the relaxed lattice structure. Scanning transmission electron microscopy (STEM) further confirms the epitaxial stabilization and "cube-on-cube" epitaxy of 3C BRO on STO substrate, along with atomically sharp surface and interface, as shown in Figure 1(e). The in-plane and out-of-plane lattice parameters obtained from STEM are also consistent with the XRD results (Figure S3).

The successful epitaxial stabilization of the 3C BRO thin film enabled us to make a standard comparison of the three members of the perovskite Ruthenate family complementing our understanding of the materials system in its entirety. The structural parameters of the perovskite Ruthenates are summarized in **Table 1**. In the bulk form, CRO and SRO have orthorhombic structures with $GdFeO_3$-type octahedral distortion, whereas 3C BRO has a simple cubic structure without any distortions. On the contrary, the $A$RuO$_3$ thin films (~30 nm) grown on the (001) STO substrates, depending on the degree of strain, can be represented as having tetragonal structures. In both cases, the overall lattice parameters and (pseudo-cubic) unit cell volumes increase as the ionic radius of the $A$-site ion increases, as expected. The pseudo-cubic unit cell volumes of the CRO, SRO, and 3C BRO thin films are 58.25, 60.23, and 66.23 Å$^3$, respectively, a striking 13.7 % increase when comparing 3C BRO with CRO. This wide tunability of the unit cell volume in the perovskite Ruthenates provides an essential opportunity to customize the structure-related physical properties. For example, as a consequence of Ba doping in bulk $Sr_{1-x}Ba_xRuO_3$, the Ru–O bond is stretched, becoming longer than the $A$–O bond in the cubic phase.[24, 25] While cation non-stoichiometry could also induce larger unit cell volume.

The change in the lattice structures of the perovskite $A$RuO$_3$ thin films naturally leads to changes in their electronic structures and consequently their magnetic properties. **Figure 2**(a) shows the real part of the optical conductivity ($\sigma_1(\omega)$) of a 3C BRO thin film obtained by spectroscopic ellipsometry. The $\sigma_1(\omega)$ of CRO and SRO thin films on (001) STO substrates are also shown for comparison.[19, 26] All of the perovskite $A$RuO$_3$ thin films show qualitatively similar $\sigma_1(\omega)$, which originate from the common Ru$^{4+}$ orbital state and its interaction with oxygen, that is, Drude absorption at low photon energy and four interband transitions, manifesting the correlated metallic nature. Note that the other polymorph phases of BRO with Ru-Ru bond exhibit entirely different $\sigma_1(\omega)$ spectra, and hence, a distinctive electronic structure.[27, 28] The $\sigma_1(\omega)$ of the perovskite $A$RuO$_3$ thin films commonly exhibit two $d$–$d$ transition peaks at 1.6–1.9 ($\alpha$) and 4.3–4.5 eV ($\beta$), and two charge transfer transition peaks at 3.0–3.3 (A) and > ~6 eV (B). These peaks ($\alpha$, $\beta$, A, and B) are routinely attributed to the optical transitions between the orbital states of Ru 4$d$ $t_{2g}$ → $t_{2g}$, and Ru 4$d$ $t_{2g}$ → $e_g$, O 2$p$ → Ru 4$d$ $t_{2g}$, and O 2$p$ → Ru 4$d$ $e_g$, respectively.[29] To quantitatively compare the optical transitions among the $A$RuO$_3$ thin films, the $\sigma_1(\omega)$ were fitted using Drude–Lorentz oscillators.[29-31] **Table 2** summarizes the values of the optical transition energy and spectral weight ($W_s \equiv \int \sigma(\omega)d\omega$), for each thin film. Among them, $W_s$ values were plotted for $A$RuO$_3$, as shown in Figure 2(b). $W_{s\alpha}$ and $W_{s\beta}$ for the $d$–$d$ transitions decreased monotonically, whereas $W_{sA}$ exhibited a maximum for the SRO. $W_{sB}$ could not be quantitatively analyzed as peak B extended beyond our experimental spectral range of 5.5 eV. The charge transfer transitions (peak A and B) are essential for describing the $p$–$d$ hybridization strength between the O 2$p$ and Ru 4$d$ orbital states[26], and therefore, the magnetic exchange interaction in an itinerant magnetic system. Hence, we estimated the relative contribution of $W_{sA}$ from the total $W_s$ ($W_{stot}$) of each thin film, as shown in Figure 2(c). The relative contribution of $W_{sA}$ to the total optical transitions was the least for the CRO and the most for the SRO and was intermediate for the 3C BRO thin films. This tendency follows the trend of the ferromagnetic transition temperature ($T_c$), which is discussed as follows (Figure 2(c)).

The epitaxially-stabilized 3C BRO thin film shows itinerant ferromagnetic behavior resembling that of the SRO thin film (Figure S4). **Figure 3** shows the magnetic properties of a 3C BRO thin film on the STO substrate. The contribution from the substrate was carefully subtracted (Figure S5). The zero-field-cooled (ZFC) and field-cooled (FC) temperature-dependent magnetization, $M(T)$, for the 3C BRO thin film are obtained along the out-of-plane and in-plane directions, as shown in Figure 3(a) and Figure S6, respectively. The 3C BRO thin film exhibits a ferromagnetic transition with a clearly discernible $T_c$ of ~48 K, as is evident from the $dM/dT$ curves for both the out-of-plane and in-plane directions (insets of Figure 3(a) and Figure S6(a)). In the bulk form, 3C BRO is ferromagnetic with $T_c = 60$ K.[32] However, under pressure, the $T_c$ decreases dramatically to 39 K above 4 GPa, a trend that is consistent with our result for the partially strained, epitaxially-stabilized 3C BRO thin film.[33] We also note a considerable $M$ remaining above $T_c$, possibly owing to magnetic fluctuation that survives up to a higher $T$. The magnetic field-dependent magnetization, $M(H)$, shows a clear hysteresis loop opening below $T_c$ (Figure 3(b) and Figure S6(b)). The saturation $M$ (0.5 $\mu_B$/Ru) of the thin film is slightly smaller than that of the bulk 3C BRO (0.8 $\mu_B$/Ru).[16] Even though we observed somewhat larger out-of-plane $M$ than in-plane $M$, the ferromagnetism in the 3C BRO thin film is more isotropic than that in SRO thin films, possibly owing to its original cubic symmetry.

The magnetic ordering of the perovskite Ruthenates thin films, that is, paramagnetism in CRO, and ferromagnetism in SRO ($T_c$ = ~150 K)[16] (Figure S4) and BRO ($T_c$ = ~50 K) (Figure 3), could be collectively understood in terms of the magnitude of the orbital hybridization. As previously discussed, the optical charge transfer transition determines the strength of the $p$–$d$ hybridization. This facilitates the exchange interaction between Ru ions connected via O ions, a result that is supported by those of a band calculation study.[31, 34] Indeed, when $T_c$ was plotted along with $W_{sA}/W_{stot}$, as shown in Figure 2(c), the same $A$-site dependency is clearly demonstrated, thereby validating the importance of the hybridization strength in understanding the itinerant magnetism in perovskite Ruthenates.

To further explore the itinerant nature of the magnetism of 3C BRO, the transport properties of the thin film were measured, as shown in **Figure 4**. The metallic $T$-dependent resistivity ($\rho(T)$) of the 3C BRO thin film is similar to that of the CRO thin film. On the other hand, it also shows an anomaly at 60 K, signifying the $T_c$, as in the case of SRO thin film (Figure S7). The resistivity values of 3C BRO are smaller than those of the reported bulk sample, probably owing to the phase purity of our thin film. However, the $\rho(T)$ was still larger than those of the SRO thin film across the experimental temperature range.[19, 26, 30] The low temperature $\rho(T)$ was further fitted using the power law $\rho(T) = \rho_0 + T^\alpha$ (where $\rho_0$ is the residual resistivity and $\alpha$ is a scaling exponent), to reveal a NFL phase with $\alpha$ = ~1.4 below ~25 K (Figure S8). In a pressure-dependent study of the bulk 3C BRO, the value of $\alpha$ was ~1.4 under the pressure of 8 GPa, which is consistent with our observation.[33]

The magneto-transport results of the 3C BRO thin film provide additional support for the existence of the NFL phase. Figure 4(b) shows the magnetic field-dependent Hall resistivity ($\rho_{xy}(H)$). The 3C BRO thin film exhibits nonlinear behavior in $\rho_{xy}$ (anomalous Hall term), which is induced by the ferromagnetic order at low temperature. Note that the normailzed $\rho_{xy}(H)$ curve qualitatively coincides with the normalized $M(H)$ curve (Figure S9), except for the hysteresis loop opening, possibly due to the low $H$-field sensitivity of the transport measurement or small $M$ of the 3C BRO thin film. The hole carrier density, $n_h$ = 5.53 × $10^{22}$ cm$^{-3}$ at 50 K, was extracted from the linear part of $\rho_{xy}$. At lower temperature of 25 and 2 K, this value decreased slightly to 4.73 × $10^{22}$ and 4.66 × $10^{22}$ cm$^{-3}$, respectively. These values are comparable to those in conventional metals, including ferromagnetic SRO thin film.[35, 36] Figure 4(c) shows the magnetoresistance (MR) at different temperatures. By scaling the MR and zero-field resistivity, we drew Kohler's plot, as shown in the inset of Figure 4(c). Based on Kohler's rule, if the carrier density of the system is robust to temperature variation, the MR curves measured at different temperatures can be scaled into a single curve. Because Kohler's rule is based on the linearity of the Boltzmann equation with the electron scattering lifetime, any deviation from the rule presents as evidence for NFL behavior.[37] As shown in the inset of Figure 4(c), the scaled MR curves violate Kohler's rule, thus supporting the NFL behavior of 3C BRO, in addition to the fact that the scaling exponent $\alpha$ deviates from 2. On the other hand, the positive MR observed in the ferromagnetic phase of 3C BRO thin film is rather unusual. While there are several mechanisms reported for the observation of positive MR in various ferromagnetic materials,[38] the small *M* and possible magnetic fluctuation above $T_c$ might be attributed to the positive MR in 3C BRO thin film.

## Conclusion

In summary, we successfully prepared epitaxially-stabilized 3C BRO thin films to complement the perovskite phases of Ruthenate thin films. The 3C BRO thin films grown under optimal conditions revealed a tetragonal structure with a four-fold symmetry, indicative of the surface epitaxial stabilization. The variation in the lattice structures within the family of perovskite *A*RuO$_3$ thin films naturally leads to changes in the electronic structures, and consequent magnetic properties. Indeed, we confirmed the strong correlation between the ferromagnetism of the perovskite Ruthenates and the orbital hybridization strength. The epitaxially-stabilized 3C BRO thin film showed itinerant ferromagnetism with a transition temperature of ~48 K. A clear magnetic hysteresis loop was presented with a saturation magnetic moment of ~0.5 $\mu_B$/Ru. Furthermore, *dc* and magneto-transport results indicate that the 3C BRO thin film behaves like a NFL at low temperatures. Our study provides a fundamental understanding of the controllability of the systematic evolution of the physical properties within cubic perovskite Ruthenate thin films.

## Experimental Section/Methods

*Thin film growth and structural characterization*: High-quality epitaxial BaRuO$_3$ (BRO) thin films were grown on atomically flat SrTiO$_3$ (STO) (001) single crystalline substrates using Pulsed Laser Epitaxy (PLE) at 750 °C. Laser (248 nm; Lightmachinery, IPEX 864) fluence of 1.0 J/cm$^2$ and a repetition rate of 5 Hz was used. The vertical distance between the target and the substrate was fixed to 65 mm. In order to successfully prepare 3C BRO, the films were grown under various oxygen partial pressure ($P(O_2)$) conditions, ranging from $30^{-1}$ to $10^{-3}$ Torr. The atomic structure, crystal orientations, and the epitaxial relation of BRO thin films High-quality using high resolution x-ray diffraction (XRD), in-plane $\varphi$-scan, and reciprocal space map (RSM). The thickness of all the BRO thin films was ~30 ± 1 nm from X-ray reflectometry (XRR). Cross-sectional transmission electron microscopy (TEM) specimens were prepared by ion milling at the LN2 temperature after conventional mechanical polishing. High-angle annular dark field scanning transmission electron microscopy (HAADF STEM) measurements were performed on Nion UltraSTEM200 operated at 200 kV. The microscope was equipped with a cold field emission gun and a corrector of third- and fifth-order aberration for sub-angstrom resolution. The collection inner and outer half-angles for HAADF STEM were 65 and 240 mrads, respectively.

*Ellipsometry*: The optical properties of the BRO thin films were investigated using spectroscopic ellipsometers (VASE and M-2000, J. A. Woollam Co.) at room temperature. The optical spectra were obtained between 0.74 and 5.5 eV for the incident angles of 65°. Two-layer model (BRO thin film on STO substrate) was sufficient for obtaining physically reasonable spectroscopic dielectric functions of BRO.

*Resistivity and magnetization measurements*: Resistivity as a function of temperature, $\rho(T)$, were measured using establishes closed-cycle refrigerator. The measurements were performed from 300 to 10 K, using Van der Pauw method with In electrodes and Au wires. Magnetotransport measurements were performed in magnetic fields up to 9 T and temperatures down to 2 K in a quantum design physical properties measurement system. $M(T)$ was measured using Magnetic Property Measurement System (MPMS, Quantum Design). The measurements were performed from 300 to 2 K under 1000 *Oe* of magnetic field along the in-plane and out-of-plane direction.

## Acknowledgements

This work was supported by Basic Science Research Programs through the National Research Foundation of Korea (NRF) (NRF-2019R1A2B5B02004546, NRF-2019R1A2C1005267 (S.A L.), and11

NRF-2020R1F1A1073076 (J-.Y.H)) and Korea Basic Science Institute (National Research Facilities and Equipment Center) grant funded by the Ministry of Education (2019R1A6C1020015). The transport measurements and analyses at ORNL were supported by the U.S. Department of Energy, Office of Science, Basic Energy Sciences, Materials Sciences and Engineering Division.

**Figures**

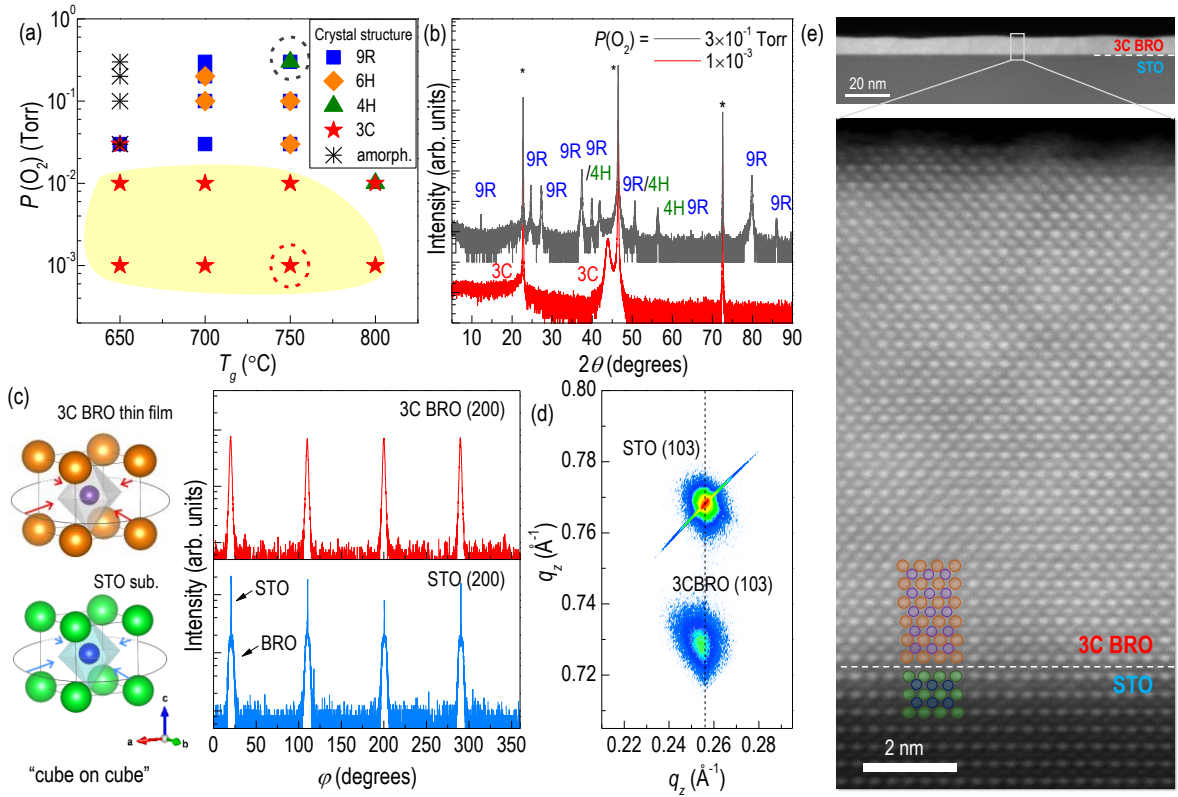

**Figure 1.** Symmetry stabilization of polymorph BaRuO$_3$ thin films on (001) SrTiO$_3$ substrates. (a) Growth window for the 3C BaRuO$_3$ thin films. With decreasing $P(O_2)$, the 3C BaRuO$_3$ thin film becomes the dominant phase. (b) XRD $\theta$–$2\theta$ scans of the BaRuO$_3$ thin films grown at $P(O_2) = 3\times10^{-1}$ and $1\times10^{-3}$ Torr, respectively. The polycrystalline BaRuO$_3$ thin film is observed at high $P(O_2)$, whereas the BaRuO$_3$ thin film clearly exhibits the (002)$_{3C}$ peak without any secondary peaks at low $P(O_2)$. (c) Crystal structure and XRD in-plane four-fold $\varphi$-scan using (200)$_{3C}$ planes for 3C BaRuO$_3$ thin film and SrTiO$_3$ (200) Bragg reflection. (d) XRD reciprocal space map of the 3C BaRuO$_3$ thin film grown around the (103) Bragg reflection of the SrTiO$_3$ substrate, which shows a partially strained thin film. (e) Low (upper figure) and high (lower figure) magnification HAADF STEM images of the 3C BaRuO$_3$ thin film on SrTiO$_3$ substrate.

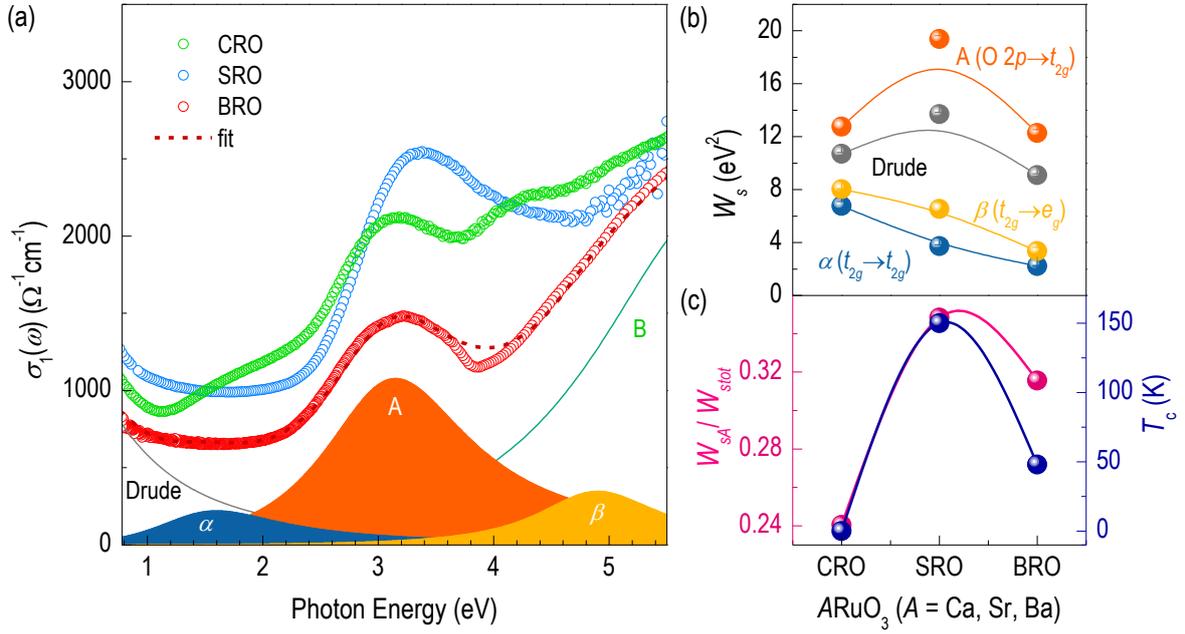

**Figure 2.** Electronic structure of the $A$RuO$_3$ thin films on SrTiO$_3$ substrate. (a) Real part of the optical conductivity, $\sigma_1(\omega)$, of the $A$RuO$_3$ thin films at room temperature. The symbols denote the experimental spectra and the dashed lines represent the results of Drude–Lorentz fitting, which are in a good agreement. The lined curves represent the deconvoluted Lorentzian peaks corresponding to each characteristic optical transition, indicating Drude, $\alpha$ ($d$–$d$ transition between Ru $t_{2g}$ states), A (charge transfer transition between O 2$p$ and Ru $t_{2g}$), $\beta$ ($d$–$d$ transition between Ru $t_{2g}$ and $e_g$ states), and B (charge transfer transition between O 2$p$ and Ru $e_g$). (b) Spectral weight ($W_s$) evolution of the Lorentz oscillators and (c) spectral weight ratio for total $W_s$ ($W_{stot}$) and charge transfer transition ($W_{sA}$) as a function of the volume of $A$RuO$_3$.

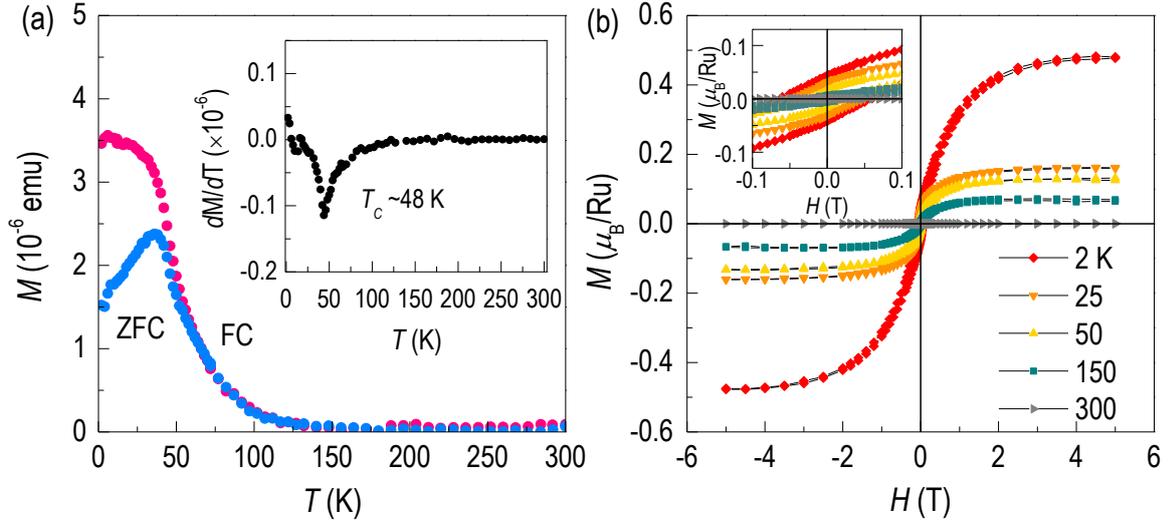

**Figure 3.** Ferromagnetic characteristic of a cubic perovskite BaRuO$_3$ thin film on a SrTiO$_3$ substrate along the out-of-plane direction. (a) $M(T)$ curves recorded at 1000 Oe. The inset presents the $dM/dT$ curve with distinct ferromagnetic $T_c$. (b) $M(H)$ hysteresis loops at different temperatures. The inset shows the low magnetic field region.

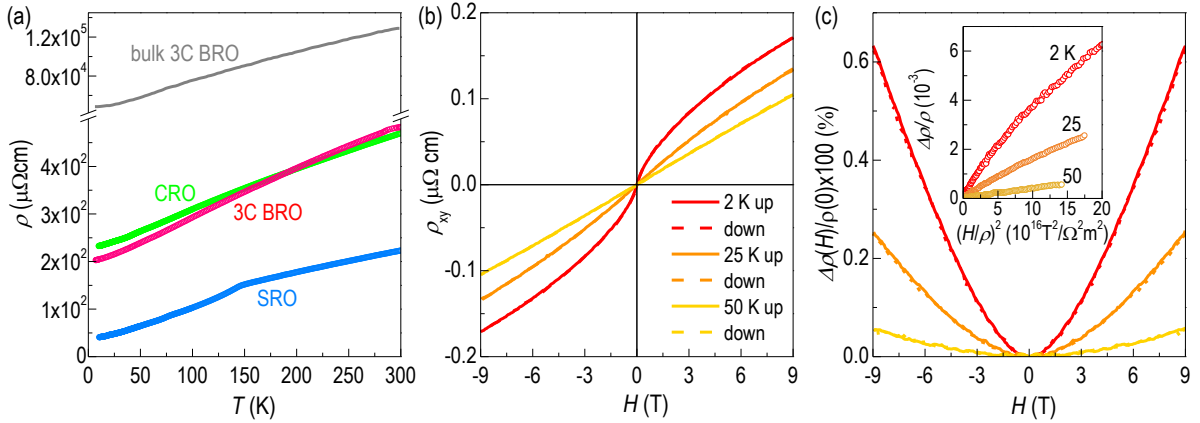

**Figure 4.** Electronic properties of a 3C BaRuO$_3$ thin film. (a) Resistivity ($\rho(T)$), (b) Hall resistivity ($\rho_{xy}(H)$) at various temperatures, and (c) magnetoresistance of a 3C BaRuO$_3$ thin film. The inset in (c) presents Kohler's plots.

**Table 1.** Comparison of $A$RuO$_3$ ($A$ = Ca, Sr, and Ba) materials in the bulk and thin film forms on (001) SrTiO$_3$ substrates. For bulk $A$RuO$_3$, the $A$-site ionic radius, crystal structure, lattice constant, and <$A$–O> bond length were obtained from the literature.[24] The information of the thin film was obtained from experiments conducted in the current study, including the symmetry-stabilized 3C BaRuO$_3$.

| | | Bulk [21] | | | | Thin Film (~30 nm on STO substrate) | | | |
|---|---|---|---|---|---|---|---|---|---|
| | $A$-site ionic radius (Å) | bulk crystal structure (Å) | pseudo-cubic lattice const. (Å) | <$A$-O> bond length (Å) | pseudo-cubic unit cell volume (Å$^3$) | strain | film structure (Å) | pseudo-tetragonal lattice const. (Å) | [$c_{pc}$/$a_{pc}$] | pseudo-cubic unit cell volume (Å$^3$) |
| CRO | 1.00 | Orthorhombic $a$ = 5.36 $b$ = 5.54 $c$ = 7.68 | $a_{pc}$ = 3.85 | 2.73 | 56.92 | Tensile 1.7 % (fully strained) | Tetragonal $a$ = $b$ = 5.52 $c$ = 7.64 | $a_{pt}$ = $b_{pt}$ = 3.905 $c_{pt}$ = 3.82 | 0.98 | 58.25 |
| SRO | 1.18 | Orthorhombic $a$ = 5.57 $b$ = 5.53 $c$ = 7.85 | $a_{pc}$ = 3.93 | 2.78 | 60.70 | Compressive 0.64 % (fully strained) | Orthorhombic $a$ = 5.60 $b$ = 5.55 $c$ = 7.81 | $a_{pt}$ = $b_{pt}$ = 3.905 $c_{pt}$ = 3.95 | 1.01 | 60.23 |
| 3C BRO | 1.36 | Cubic $a$ = $b$ = $c$ = 4.00 | | 2.83 | 64.00 | Compressive 2.46 % (partially strained) | Tetragonal $a_{pc}$ = $b_{pc}$ = 3.99 $c_{pc}$ = 4.16 | | 1.03 | 66.23 |

**Table 2.** Parameters of the deconvoluted Lorentzian peaks including the optical transition energy and spectral weight for the $A$RuO$_3$ thin films.

| | Optical conductivity | | | | | | | | |
|---|---|---|---|---|---|---|---|---|---|
| | Optical transition energy (eV) | | | | | Spectra weight (eV$^2$) | | | |
| | $\alpha$ | A | $\beta$ | B | Drude | $W_{s\alpha}$ | $W_{sA}$ | $W_{s\beta}$ | $W_{sB}$ |
| CRO | 1.89 | 3.03 | 4.27 | 5.94 | 10.70 | 6.76 | 12.76 | 8.02 | 56.45 |
| SRO | 1.70 | 3.25 | 4.20 | 6.00 | 13.71 | 3.73 | 19.38 | 6.53 | 48.60 |
| 3C BRO | 1.60 | 3.11 | 4.45 | 6.00 | 9.10 | 2.23 | 12.28 | 3.36 | 47.16 |

Supporting Information

# Epitaxial stabilization of metastable 3C $BaRuO_3$ thin film with ferromagnetic non-Fermi liquid phase

Sang A Lee[1,2], Jong Mok Ok[3], Jegon Lee[1], Jae-Yeol Hwang[2], Sangmon Yoon[3], Se-Jeong Park[4], Sehwan Song[5], Jong-Seong Bae[6], Sungkyun Park[5], Ho Nyung Lee[3], and Woo Seok Choi[1*]

[1]Department of Physics, Sungkyunkwan University, Suwon 16419, Korea

[2]Department of Physics, Pukyong National University, Busan 48513, Korea

[3]Materials Science and Technology Division, Oak Ridge National Laboratory, Oak Ridge, TN 37831, U.S.A.

[4]Application Group, Korea I.T.S. Co. Ltd., Seoul 06373, Korea

[5]Department of Physics, Pusan National University, Busan 46241, Korea

[6]Busan Center, Korea Basic Science Institute, Busan 46742, Korea

* e-mail: choiws@skku.edu

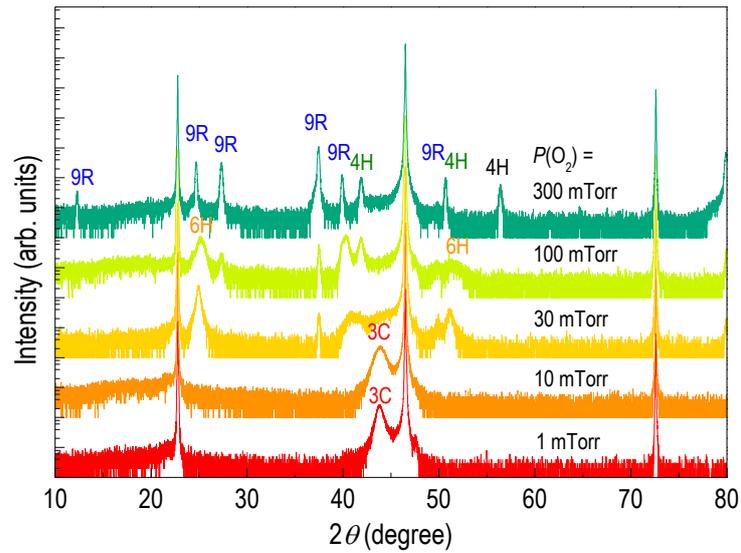

**Figure S1.** XRD $\theta$-$2\theta$ scans of the BaRuO$_3$ thin films on SrTiO$_3$ (001) substrate grown at various oxygen partial pressure ($P$(O$_2$)) from 300 to 1 mTorr at 750 °C.

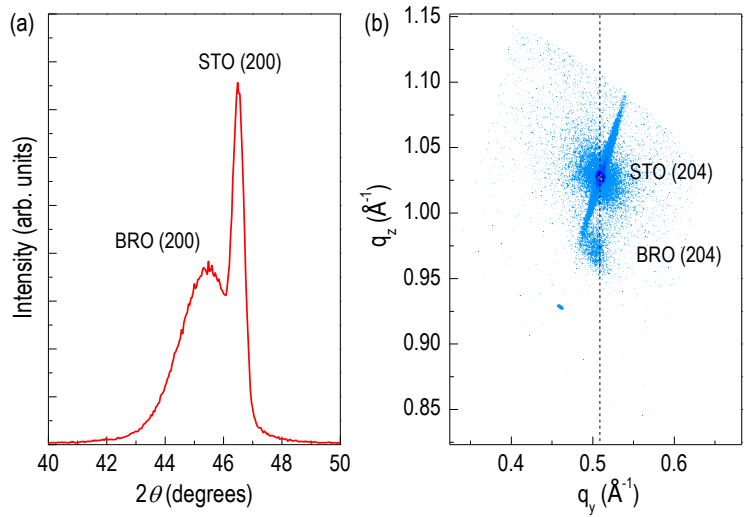

**Figure S2.** (a) XRD in-plane $\theta$-$2\theta$ scans around the 3C BaRuO$_3$ (200) thin film and SrTiO$_3$ (200) Bragg reflection. (b) Wide range XRD reciprocal space map of the BaRuO$_3$ thin film grown around the (204) Bragg reflection of the SrTiO$_3$ substrate.

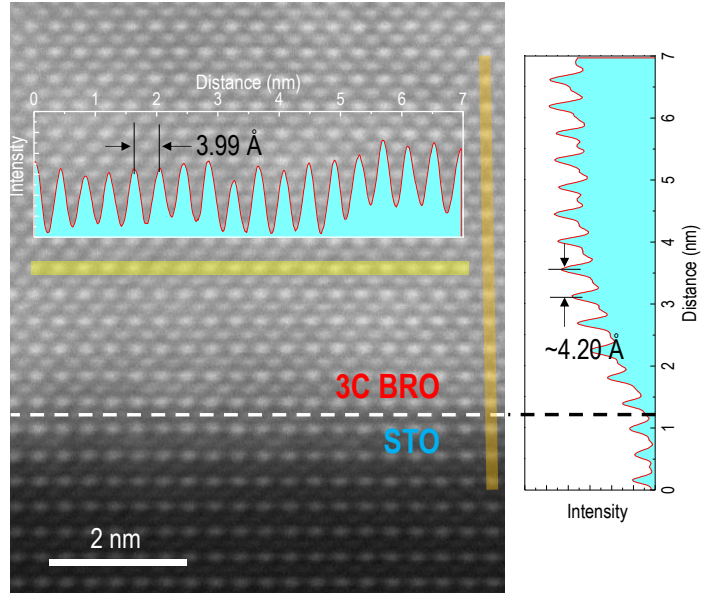

**Figure S3.** In-plane and out-of-plane lattice parameters obtained from the STEM image of the 3C BaRuO$_3$ thin film on SrTiO$_3$ substrate.

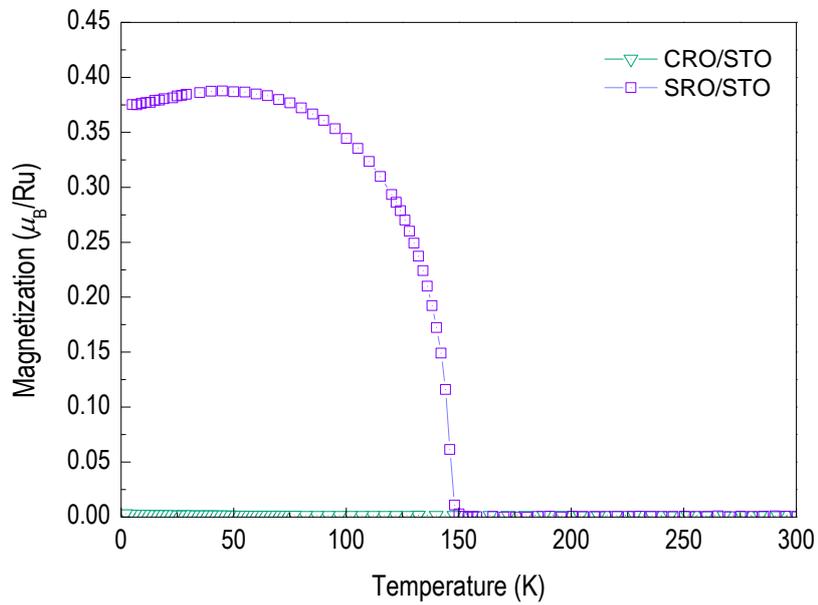

**Figure S4.** $M(T)$ curves for the CaRuO$_3$ and SrRuO$_3$ thin films at 100 Oe. The CaRuO$_3$ and SrRuO$_3$ thin films show paramagnetic and ferromagnetic (Curie temperature of ~150 K) behaviors, respectively.

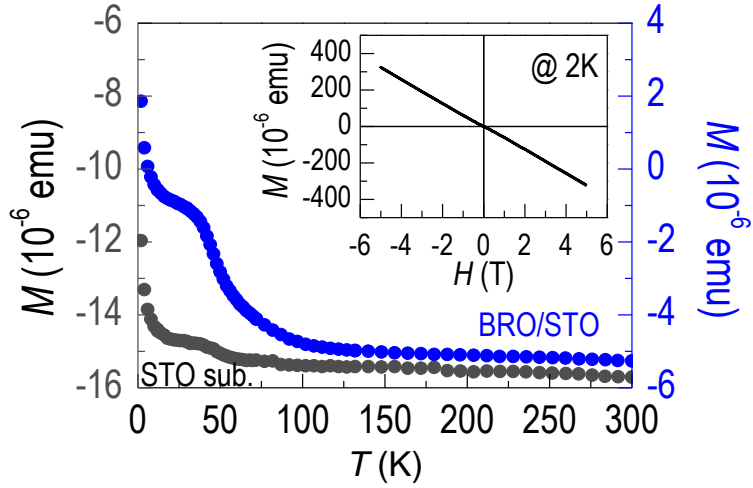

**Figure S5.** M(T) curves for a SrTiO$_3$ substrate and BaRuO$_3$ thin film on SrTiO$_3$ substrate at 1000 Oe. The insets is $M(H)$ curves for SrTiO$_3$ substrate, respectively, at 2 K.

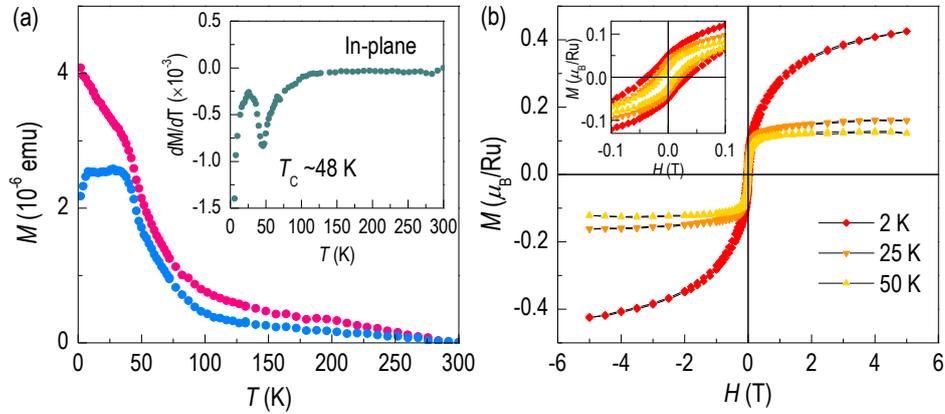

**Figure S6.** Ferromagnetic characteristic of a cubic perovskite BaRuO$_3$ thin film on a SrTiO$_3$ substrate. $M(T)$ curves for a 3C BaRuO$_3$ thin film along the (a) in-plane directions at 1000 Oe. The insets in (a) is the $dM/dT$ curves. $M(H)$ hysteresis loops for the 3C BaRuO$_3$ thin film at different temperatures along the (b) in-plane. The insets in the figure show the lower magnetic field region.

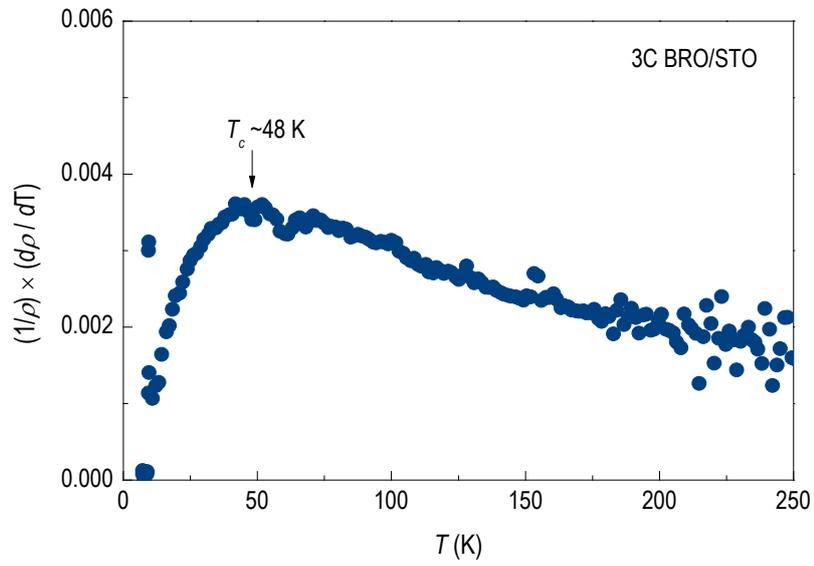

**Figure S7.** The $(1/\rho) \times d\rho/dT$ curve of the 3C BaRuO$_3$ thin film.

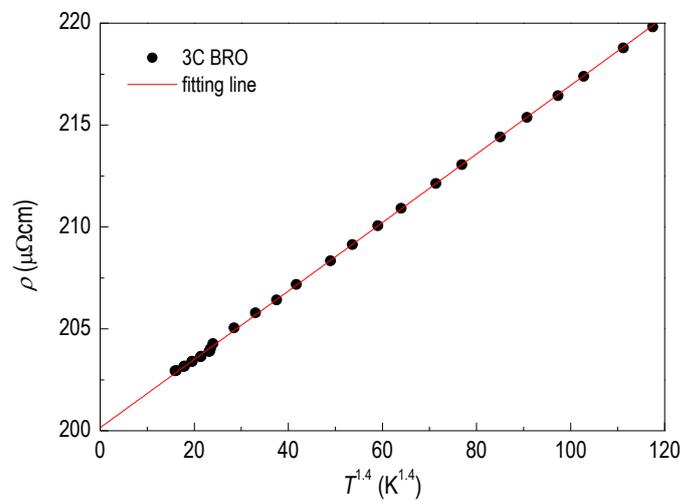

**Figure S8.** The $\rho$ versus $T^{\alpha}$ plot of the 3C BaRuO$_3$ thin film at low temperature.

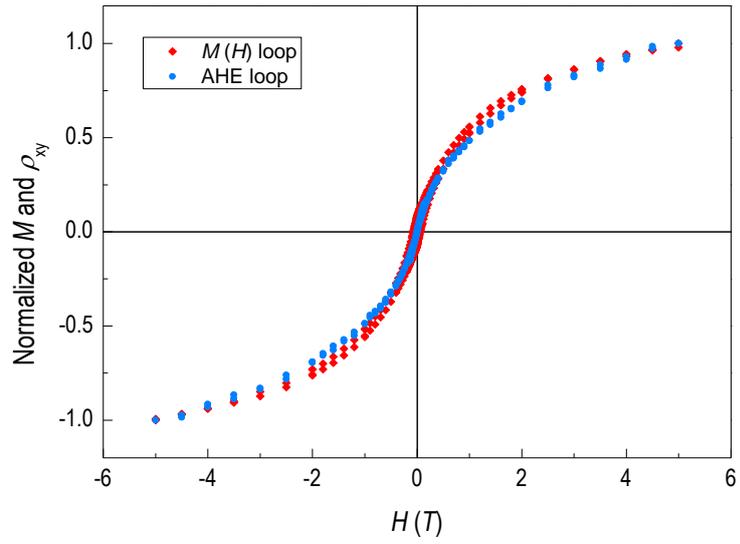

**Figure S9.** Hysteresis loop of 3C BRO thin film for normalized $M$ and $\rho_{xy}$ versus $H$.